\documentclass[10pt, letterpaper]{IEEEtran}
\usepackage{todonotes}
\usepackage[nocompress]{cite}
\usepackage{url}
\usepackage{enumitem}
\usepackage{epstopdf}
\PrependGraphicsExtensions{.tif, .tiff}
\usepackage{graphicx}
\usepackage{caption}
\usepackage{amsmath}
\usepackage{tabu}
\usepackage{multirow}
\usepackage{adjustbox}
\usepackage{mathptmx}
\usepackage{amsfonts}
\usepackage{algpseudocode}
\usepackage{soul}
\usepackage[bookmarks=false]{hyperref}
\usepackage{footnote}
\usepackage{lipsum}
\usepackage[ruled,vlined, linesnumbered]{algorithm2e}
\usepackage{subcaption}
\usepackage{listings}
\usepackage{tikz}
\usetikzlibrary{positioning}
\soulregister\cite7
\soulregister\ref7
\soulregister\pageref7
\soulregister\secref7
\soulregister\tabref7

\definecolor{ao}{rgb}{0.0, 0.5, 0.0}
\definecolor{db}{rgb}{0.2, 0.2, 0.6}
\definecolor{cadmiumgreen}{rgb}{0.0, 0.42, 0.24}

\newcommand{\tabref}[1]{Table~\ref{#1}}

\newcommand{\secref}[1]{Section~\ref{#1}}

\renewcommand{\algref}[1]{Algorithm~\ref{#1}}
\SetAlFnt{\footnotesize}

\newcommand\copyrighttext{%
  \footnotesize \copyright 2020 IEEE. Personal use of this material is permitted. Permission from IEEE must be obtained for all other uses, in any current or future media, including
reprinting/republishing this material for advertising or promotional purposes, creating new
collective works, for resale or redistribution to servers or lists, or reuse of any copyrighted
component of this work in other works.}
\newcommand\copyrightnotice{%
\begin{tikzpicture}[remember picture,overlay]
\node[anchor=south,yshift=10pt] at (current page.south) {\fbox{\parbox{\dimexpr\textwidth-\fboxsep-\fboxrule\relax}{\copyrighttext}}};
\end{tikzpicture}%
}

\begin{document}
\title{Extending SLURM for Dynamic Resource-Aware Adaptive Batch Scheduling}
\author{\IEEEauthorblockN{Mohak Chadha\IEEEauthorrefmark{1}, Jophin John\IEEEauthorrefmark{2}, Michael Gerndt\IEEEauthorrefmark{3}}
\IEEEauthorblockA{\\Chair of Computer Architecture and Parallel Systems, Technische Universit{\"a}t M{\"u}nchen \\
Garching (near Munich), Germany} \\
Email: mohak.chadha@tum.de, john@in.tum.de, gerndt@in.tum.de}

\maketitle
\copyrightnotice


\begin{abstract}

With the growing constraints on power budget and increasing hardware failure rates, the operation of future exascale systems faces several challenges. Towards this, resource awareness and adaptivity by enabling malleable jobs has been actively researched in the HPC community. Malleable jobs can change their computing resources at runtime and can significantly improve HPC system performance. However, due to the rigid nature of popular parallel programming paradigms such as MPI and lack of support for dynamic resource management in batch systems, malleable jobs have been largely unrealized. In this paper, we extend the SLURM batch system to support the execution and batch scheduling of malleable jobs. The malleable applications are written using a new adaptive parallel paradigm called Invasive MPI which extends the MPI standard to support resource-adaptivity at runtime. We propose two malleable job scheduling strategies to support performance-aware and power-aware dynamic reconfiguration decisions at runtime. We implement the strategies in SLURM and evaluate them on a production HPC system. Results for our performance-aware scheduling strategy show improvements in makespan, average system utilization, average response, and waiting times as compared to other scheduling strategies. Moreover, we demonstrate dynamic power corridor management using our power-aware strategy.



\end{abstract}


\IEEEpeerreviewmaketitle

\section{Introduction}
\label{sec:intro}


A critical component in modern HPC systems is a middleware called Resource and Job Management System (RJMS) software. RJMS is responsible for efficiently distributing the computing resources among the users of an HPC system and mapping the submitted jobs to the underlying hardware. It consists of three primary subsystems namely, Resource Management (RM), Job Management (JM) and Scheduler~\cite{georgiou2010contributions}. The combination of the RM subsystem along with the scheduler constitutes the batch system.

As we move towards future exascale system their operation faces several crucial challenges such as energy efficiency and fault tolerance~\cite{khaleel2011scientific}. For example, the power consumption of supercomputing centers must be bounded within a power corridor, according to the contract with energy companies. To allow the control of the overall power consumption in the grid, the centers can agree to a dynamic power corridor management by the power supplier~\cite{chen2014data}. For this they get a reduction in energy costs. Furthermore, HPC applications are becoming more dynamic. For instance, scientific applications that utilize Adaptive Mesh Refinement (AMR)~\cite{plewa2005adaptive, mo2017large} techniques change their resource requirements at runtime due to varying computational phases based upon refinement or coarsening of meshes. A solution to overcome these challenges and efficiently utilize the system components is resource awareness and adaptivity. A method to achieve adaptivity in HPC systems is by enabling malleable jobs.



A job is said to be malleable if it can adapt to resource changes triggered by the batch system at runtime~\cite{feitelson1996toward}. These resource changes can either increase (expand operation) or reduce (shrink operation) the number of processors. Malleable jobs have shown to significantly improve the performance of a batch system in terms of both system and user-centric metrics such as system utilization and average response time~\cite{gupta2014towards, utrera2012job}. Enabling malleable jobs in HPC systems requires an adaptive parallel runtime system and an adaptive batch system. An adaptive batch system requires an adaptive job scheduler and a dynamic resource manager. The adaptive job scheduler is responsible for deciding which running malleable jobs to expand or shrink depending upon the current job queue and the job scheduling strategy. The primary function of the dynamic resource manager is to enforce these decisions to the running jobs while maintaining a consistent system state.


To facilitate the development of resource-elastic applications several adaptive parallel paradigms have been developed. These include OmpSs~\cite{duran2011ompss}, particularly for shared memory systems, Charm++~\cite{kale2002malleable}, and Adaptive Message Passing Interface  (AMPI)~\cite{huang2003adaptive} for distributed memory systems. Charm++ supports a message driven execution model and allows programmers to define objects, i.e., units of work/data. It supports resource adaptation operations using task migration, checkpoint-restart, and Linux shared memory~\cite{gupta2014towards}. However, this mechanism is not transparent to the developer and the application needs to be rewritten using the programming model. AMPI builds on top of Charm++ and utilizes over-subscription of virtual MPI processes to the same CPU core to support malleability. A virtual MPI process (rank) in AMPI is a user-level thread encapsulated into a Charm++ object. 

Charm++ and AMPI do not follow the current MPI execution model of processes with private address spaces and no over-subscription~\cite{mpi_standard}. Towards this, we utilize the Invasive MPI (iMPI) library~\cite{compres2016infrastructure} which extends the MPI standard using four routines to support malleability of distributed applications at runtime. As a result, programmers can directly reuse the general structure and computational blocks of preexisting MPI applications for writing malleable jobs. 



Traditionally, batch systems in most RJMS software support the execution of only rigid jobs, i.e., the number of processors allocated remain fixed during the entire job duration~\cite{feitelson1996toward}. An example of this is SLURM~\cite{yoo2003slurm} which is an open-source, scalable workload manager installed in six of the top ten supercomputers in the world. Therefore, to support expand/shrink operations and the management of different job types current batch systems need to be extended. In this paper, we extend the batch system in SLURM to support the combined scheduling of rigid MPI and malleable iMPI based applications. We present two novel job scheduling strategies to support performance-aware and power-aware dynamic reconfiguration decisions. The performance-aware strategy utilizes the heuristic criterion \emph{MPI to Compute Time (MTCT)} ratio of running jobs for efficient expand/shrink decisions. The power-aware strategy supports dynamic power corridor management for HPC systems. If the power consumption of the system violates the limits of the power corridor, then appropriate node redistribution or scheduling decisions are taken to reinforce it.  
We evaluate and compare the performance of our performance-aware strategy against other state-of-the-art strategies wrt system and user-centric metrics by utilizing a workload with varying number of rigid and malleable jobs. Furthermore, we demonstrate the effectiveness of our power-aware scheduling strategy for maintaining the system level dynamic power corridor.


Towards energy efficiency and efficient dynamic resource management, our key contributions are:
\begin{itemize}
    \item We extend the SLURM batch system to support dynamic reconfiguration operations for malleable applications.
    \item We implement and evaluate performance-aware and power-aware job scheduling strategies on a production HPC system.
    \item We quantify and analyze the overhead for expand and shrink operations in our infrastructure.
\end{itemize}



The rest of the paper is structured as follows. Section~\ref{sec:impi} gives a brief overview of the iMPI library.  In Section~\ref{sec:related_work}, the existing techniques and frameworks for scheduling adaptive applications are described. Section~\ref{sec:adaptive_resource_management} outlines the extensions to SLURM to support dynamic resource management. In Section~\ref{sec:adaptive_job_scheduling}, the implemented job scheduling strategies are described.  Section~\ref{sec:exp_results} presents the results of our batch system. Finally, Section~\ref{sec:conclusion} concludes the paper and presents an outlook. 




\section{Background}
\label{sec:impi}


\lstset{language=C++,
                basicstyle=\ttfamily,
                keywordstyle=\color{blue}\ttfamily,
                stringstyle=\color{red}\ttfamily,
                commentstyle=\color{ao}\ttfamily,
                morecomment=[l][\color{magenta}]{\#},
                numbers=left,
                xleftmargin=2.6em,
                framexleftmargin=2.7em
}
\begin{lstlisting}[float, floatplacement=t, language=C++, frame=single, caption={Structure of a simple iMPI application.},label={lst:adapt_init}, captionpos=b,  basicstyle=\ttfamily\tiny, belowskip=-1.5 \baselineskip]
MPI_Init_adapt(...,local_status)
//Initialization Block
if local_status = newly_created_process {
    MPI_Comm_adapt_begin(...);
    //Redistrbute 
    MPI_Comm_adapt_commit();
}else{
    //preexisting processes
    phase_index = 0;
}
//Begin Elastic Block 1
if(phase_index == 0)
{
    while (elastic_block_condition){
        MPI_Probe_adapt(operation, ...);
        if (operation == resource_adaptation) {
            MPI_Comm_adapt_begin(...);
            //Redistrbute 
            MPI_Comm_adapt_commit();
        }
        iteration_count++;
        //Do computation
    }
    phase_index++;
}
//End Elastic Block 1
...
//Begin Elastic Block n
if(phase_index == n)
{
...
}
// End elastic block n
//Finalization block
...
\end{lstlisting}

The de-facto standard for programming distributed memory HPC systems is MPI. Although dynamic process support was added to the MPI standard in version 2.0 through the operations \texttt{MPI\_COMM\_SPAWN} and \texttt{MPI\_COMM\_SPAWN\_MULTIPLE}, it is rarely used due to several limitations such as high-performance overhead. Towards this, Compr\'es et al.~\cite{compres2016infrastructure} propose four routines to support dynamic process management in MPI applications. The routines are designed with latency hiding, minimal collective latency and ease of programming in mind. Furthermore, they enable ease of efficient integration with resource mangers and allow efficient implementation in MPI communication libraries.


The routines include \texttt{MPI\-\_\-Init\_adapt}, \texttt{MPI\-\_\-Probe\_\-adapt}, \texttt{MPI\-\_\-Comm\_\-adapt\_\-begin} and \texttt{MPI\-\_\-Comm\_\-adapt\_\-commit}. Listing~\ref{lst:adapt_init} shows the structure of a simple malleable application written using the four operations. In the beginning, the MPI processes of an iMPI application are initialized by calling the proposed new initialization routine (Line 1). The routine contains a \texttt{local\_status} parameter which is used for distinguishing between preexisting and newly created processes. Adaptations are performed by creating adaptation windows by using the \texttt{adapt\_begin} and \texttt{adapt\_commit} routines. In the case of newly created processes the adaptation window is started immediately (Line~3-7). On the other hand, the preexisting processes continuously check for adaptation instructions from the resource manager using the \texttt{probe\_adapt} routine (Line 15) and start the adaptation window only if they are received (Line 16-17). The adaptation starts once all the processes are at the adaptation window and the application reaches a safe location, i.e., at the beginning or end of a computation loop or phase. The application described in Listing~\ref{lst:adapt_init} is logically divided into elastic blocks where resource distribution is possible, to create suitable entry points for joining processes. The variable \texttt{phase\_index} is used to identify these entry points. After this point, data can be distributed among the new processes with the help of the helper communicators present in the \texttt{adapt\_begin} routine (Line 5, 18). Implementation of data distribution schemes is application specific and hence is the responsibility of the developer. After the adaptation completes, the global communicator \texttt{MPI\_COMM\_WORLD} is modified permanently (Line 6, 19). Following this, the application resumes its computations. In this paper, we utilize an extended version of the MPICH library~\cite{mpich} (version $3.2$) which contains the proposed routines, for writing malleable jobs and integration with SLURM~\cite{compres2016infrastructure}.

\section{Related Work}
\label{sec:related_work}
Strategies for efficient resource management and job scheduling of malleable applications have been extensively studied in the literature. However, most of the papers utilize the equipartitioning strategy~\cite{sun2011fair, blazewicz2001approximation} and are evaluated using simulations~\cite{incentive}. In contrast to the above approaches, some prototypes which combine a dynamic resource manager along with an adaptive job scheduler have been developed. Utrera et al.~\cite{utrera2012job} propose a FCFS-malleable job scheduling strategy based on the principal of virtual malleability (VM). In VM, the original number of processes are preserved, and the job is allowed to adapt to changes in the number of CPUs at runtime. The authors show that for a set of only malleable jobs, the proposed job scheduling policy leads to a 31\% improvement in average response time as compared to the widely used EASY backfilling strategy.

Prabhakaran et al.~\cite{prabhakaran2015batch} extend the Torque/Maui batch system to support dynamic reconfiguration operations for malleable jobs. They propose a Dependency-based Expand Shrink (DBES) scheduling algorithm which is capable of scheduling a combined set of rigid, evolving and malleable jobs. The algorithm utilizes dynamic fairness policies~\cite{prabhakaran2014batch}, analysis of job and resource dependencies, and 
backfilling strategy for efficient scheduling of jobs. This is the only adaptive batch system in literature for distributed memory applications and thus similar to our proposed work here. However, there are several differences. First, in~\cite{prabhakaran2015batch} malleability of applications is achieved through Charm++ runtime while we utilize iMPI. Second, we extend the batch system in SLURM while they use Torque/Maui. Third, their proposed DBES scheduling strategy does not account for performance of the application, while in our approach we use the MTCT ratio for performance-aware dynamic reconfiguration decisions. Fourth, we account for constraints on the number of nodes for dynamic reconfiguration decisions. Fifth, power-aware scheduling of malleable jobs is not discussed.




Several techniques which utilize dynamic voltage and frequency scaling~\cite{power_stat_app, ptfmodel}, software clock modulation~\cite{ddcm} and power-capping to reduce the energy consumption of HPC systems have been developed. These techniques have also been integrated into the development of power-aware job scheduling strategies for overall system savings. Sun et al.~\cite{research_power_aware} propose two scheduling policies based on intelligent backfilling and adaptive powering down of idle nodes to decrease total system power. Bodas et al.~\cite{poweraware} develop a power-aware scheduling plugin for SLURM that implements a uniform frequency mechanism, monitors power consumption and distributes a power-budget to each job. The scheduler ensures that the system operates within a certain power corridor. In contrast to previous works, we utilize adaptation operations for power-aware scheduling and dynamic power corridor management.




\begin{figure}[t]
\centering
\includegraphics[width=\columnwidth]{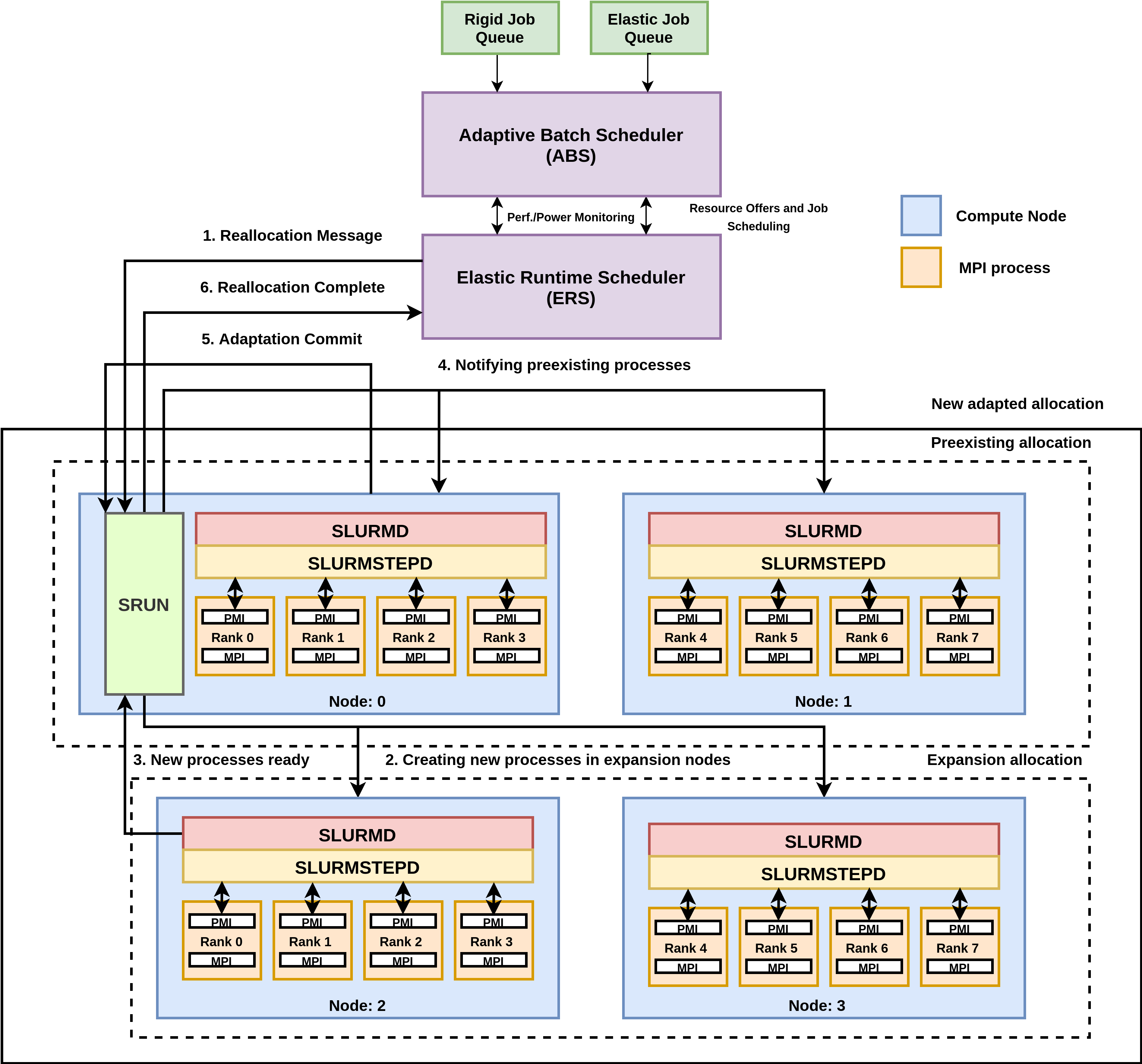}
\captionsetup{justification=centering}
\caption{Overview of interactions between different SLURM binaries during resource-adaptation operations.}
\label{fig:slurm_extension}
\end{figure}

\section{Dynamic Resource Management Infrastructure}
\label{sec:adaptive_resource_management}
In this section, we first present an overview of SLURM's architecture and the workflow for running batch jobs. Following this, we describe our extensions to SLURM for enabling dynamic resource management, in order to support scheduling, and, adaptation operations for malleable batch jobs.



\subsection{Overview of SLURM}
\label{sec:slurm_arch}
SLURM~\cite{yoo2003slurm} is an open-source, scalable, and fault-tolerant RJMS software currently being developed by SchedMD. It is a collection of several binaries and information utilities that provide different functionalities such as monitoring the machine status and partition info, submission and canceling of jobs, monitoring job queues, etc. The binaries are designed to be highly scalable and have a threaded design. Some of the important SLURM binaries include the SLURM controller \texttt{SLURMCTLD}, node daemons \texttt{SLURMD}, step daemons \texttt{SLURMSTEPD} and the interactive parallel job launcher \texttt{SRUN}. For communication between binaries and information utilities, SLURM utilizes Remote Procedure Calls (RPCs). RPCs are an important technique for building scalable distributed, client-server based applications. RPC request messages between binaries and information utilities are sent using TCP/IP.
Users can submit a job through the \texttt{sbatch} command by specifying several job-specific options such as required number of nodes, wall clock time, partition etc. in a job script. Each job script also contains at least one job step, i.e., a \texttt{srun} command for launching the application (see Listing~\ref{lst:sbatch_options}). After submission, the job is added to the priority-ordered job queue maintained by the SLURM controller. By default, the jobs are ordered wrt their arrival time. \texttt{SLURMCTLD} is the only centralized component in SLURM and is responsible for monitoring the state of each compute node and allocation of nodes to jobs. If the requested resources are available and the job has a high enough priority, the controller notifies the \texttt{SLURMD} daemon of the first allocated node to initiate the job and session managers.

The node daemons run on each compute node and periodically communicate with the controller to exchange node and job status information. On successful acknowledgement to the controller by the node daemon, the job step is initiated on the first allocated compute node. Following this, the running \texttt{SRUN} instance notifies the appropriate node daemons of the allocation to launch \texttt{SLURMSTEPD} daemons. The step daemons are responsible for launching and interacting with node-local processes of a parallel application via the Process Management Interface (PMI). Figure~\ref{fig:slurm_extension} shows the position of the different SLURM binaries in a HPC system and MPI and PMI libraries linked to the MPI processes.



\subsection{Extensions to SLURM}
\label{sec:overview_extensions}

\lstset{ %
    basicstyle=\ttfamily\footnotesize,
    commentstyle=\color{blue}\ttfamily,
    frame=single,
    keywordstyle=\color{green}\ttfamily,
    language=Bash,
    showstringspaces=false,
    morestring=[s][\color{Gray}]{<}{>},
    morestring=[s][\color{OrangeRed}]{\ -}{\ },
    morestring=[s][\color{OrangeRed}]{*}{\ },
    morestring=[s][\color{OrangeRed}]{|}{\ },
    morestring=[s][\color{OrangeRed}]{\&}{\ },
}

\begin{lstlisting}[float,floatplacement=t, language=Bash, frame=single, caption={Sample batch script with extented options for \texttt{sbatch}.},label={lst:sbatch_options}, captionpos=b,  basicstyle=\ttfamily\tiny, belowskip=-0.8 \baselineskip]
#SBATCH --job-name sample_job_script
#SBATCH --time=00:15:00
#SBATCH --nodes=1
#SBATCH --ntasks-per-node=48
#SBATCH --min-nodes-invasic=1
#SBATCH --max-nodes-invasic=5
#SBATCH --min-power=100 #in Watts
#SBATCH --max-power=200 #in Watts
#SBATCH --node-constraints="odd"
#possible values pof2, even, ncube, odd, #none

srun test_app #job step 1
\end{lstlisting}

Compr\'es et al.~\cite{compres2016infrastructure} present an early prototype for SLURM to support adaptation operations for interactive iMPI applications. In this paper, we extend and consolidate their work to support combined resource-aware batch scheduling for rigid MPI and malleable iMPI applications. 


The default \texttt{SLURMCTLD} (see Section~\ref{sec:slurm_arch}) is extended and replaced by two components, i.e., the Elastic Runtime Scheduler (ERS) and the Adaptive Batch Scheduler (ABS) as shown in Figure~\ref{fig:slurm_extension}. In addition to the traditional functions of the controller, the ERS is responsible for managing expand/shrink operations for malleable batch jobs (see Section~\ref{sec:expand_shrink_jobs}), along with runtime performance and power measurement as described in Section~\ref{sec:perf_measurement}. The ABS is essentially a SLURM scheduling plugin and is responsible for efficient batch scheduling and dynamic reconfiguration decisions for running jobs. On startup, ABS is dynamically loaded and started in a separate thread by the ERS. It maintains two separate priority-ordered queues, i.e., rigid and elastic for storing user-submitted jobs as shown in Figure~\ref{fig:slurm_extension}. In order to allow users to initiate and submit jobs to the two queues, we extend the \texttt{sbatch} command line utility. A sample batch script containing the added options is shown in Listing~\ref{lst:sbatch_options}. For submitting jobs to the elastic job queue, the user must specify the minimum and maximum nodes required for the application using \texttt{--min-nodes-invasic} and \texttt{--max-nodes-invasic} parameters. This is done because the users have the maximum knowledge about their applications and its requirements. Furthermore, our current infrastructure only supports adaptation operations at a node level granularity. To this end, any expand operation will not allocate more than the specified number of maximum nodes to the job. Similarly, any shrink operation will not deallocate the number of allocated nodes to less than the specified number of minimum nodes to the job. If the above two mentioned options are not specified then the application is added to the rigid job queue. The submitted jobs are assigned priorities wrt their arrival time. 

The users can also specify constraints on the allocated number of nodes for reconfiguration decisions using the \texttt{--node-constraints} option. The supported options are power-of-two, even, odd and cubic number of nodes. This is done to support adaptation of applications that might have different constraints on the number of processes (in our case the number of nodes) for their execution. For instance, the application Lulesh~\cite{LULESH:versions} requires a cubic number of processes for execution. The usage of these options is later discussed in Section~\ref{sec:perf_aware_scheduling}. Apart from this, the user can also specify an estimate for the minimum and maximum power required by the job per node, which is used for power-aware reconfiguration decisions and described in Section~\ref{sec:power_aware_scheduling}.

The ABS has a global knowledge about the available system resources and closely interacts with the ERS for requesting performance and power data, and communicating expand/shrink decisions for running jobs. The interaction between the scheduling loop of the ABS and the ERS is event-triggered and occurs (i) after every \texttt{SchedulerTick} seconds, set in the SLURM configuration file, (ii) on submission of a new job to either rigid or elastic job queue, or (iii) on completion of a job. 





\subsection{Expand/Shrink for Malleable Batch Jobs}
\label{sec:expand_shrink_jobs}
After the ABS decides to expand/shrink a running malleable job, it is the responsibility of the ERS to enforce the operation, while maintaining a consistent system state. Towards this, an expand/shrink reallocation handler was developed. The reallocation handler is responsible for changing the context of the running job step, launching new processes for an expand operation, and destroying the preexisting processes for a shrink operation. The handler runs in a separate thread inside the \texttt{SRUN} instance of a batch job. For communication with the handler, the ERS requires a port for communication with the particular thread and the hostname of the compute node with the running job step. The hostname is always the first node of the allocation as described in Section~\ref{sec:slurm_arch}. Before the application starts execution, the \texttt{SRUN} instance assigns a unique port number to the particular thread which is communicated to the ERS via a modified RPC request message (see Section~\ref{sec:slurm_arch}).

Runtime reconfiguration of a malleable application is a six-step process, each of is are shown in Figure~\ref{fig:slurm_extension}. In the first step, a reallocation message is generated by the ERS and sent to the \texttt{SRUN} instance, which invokes the developed handler. If the adaptation operation is an expansion, then \texttt{SRUN} notifies \texttt{SLURMD} daemons of all expansion nodes to launch the required number of processes. The \texttt{SLURMD} daemons in the expansion nodes notify \texttt{SRUN} after the newly created processes are ready at the beginning of the adaptation window (see Section~\ref{sec:impi}). In the case of a shrink operation, \texttt{SRUN} sends instructions to the \texttt{SLURMD} daemons of nodes which are to be retreated from. Following this, each \texttt{SLURMD} daemon updates its local \texttt{MPI\_Probe\_adapt} metadata. During both expand and shrink operations, the state of the job is changed to \texttt{ADAPTING} from \texttt{RUNNING}. After completion of the adaptation operation, the leader node notifies \texttt{SRUN}. Following this, \texttt{SRUN} notifies the ERS that the adaptation was completed by sending a reallocation complete message. Finally, ERS sends \texttt{SRUN} the updated job credentials and updates the job state back to \texttt{RUNNING}. It is important to note that the reallocation message must always be either a pure expand or shrink operation, i.e., mixed adaptation operations are not supported. Furthermore, our infrastructure supports simultaneous expand/shrink operations for different malleable jobs. However, the node running the \texttt{SRUN} instance must always be part of the adaptation operation and cannot be migrated.




\begin{figure}[t]
\centering
\includegraphics[width=0.65\columnwidth]{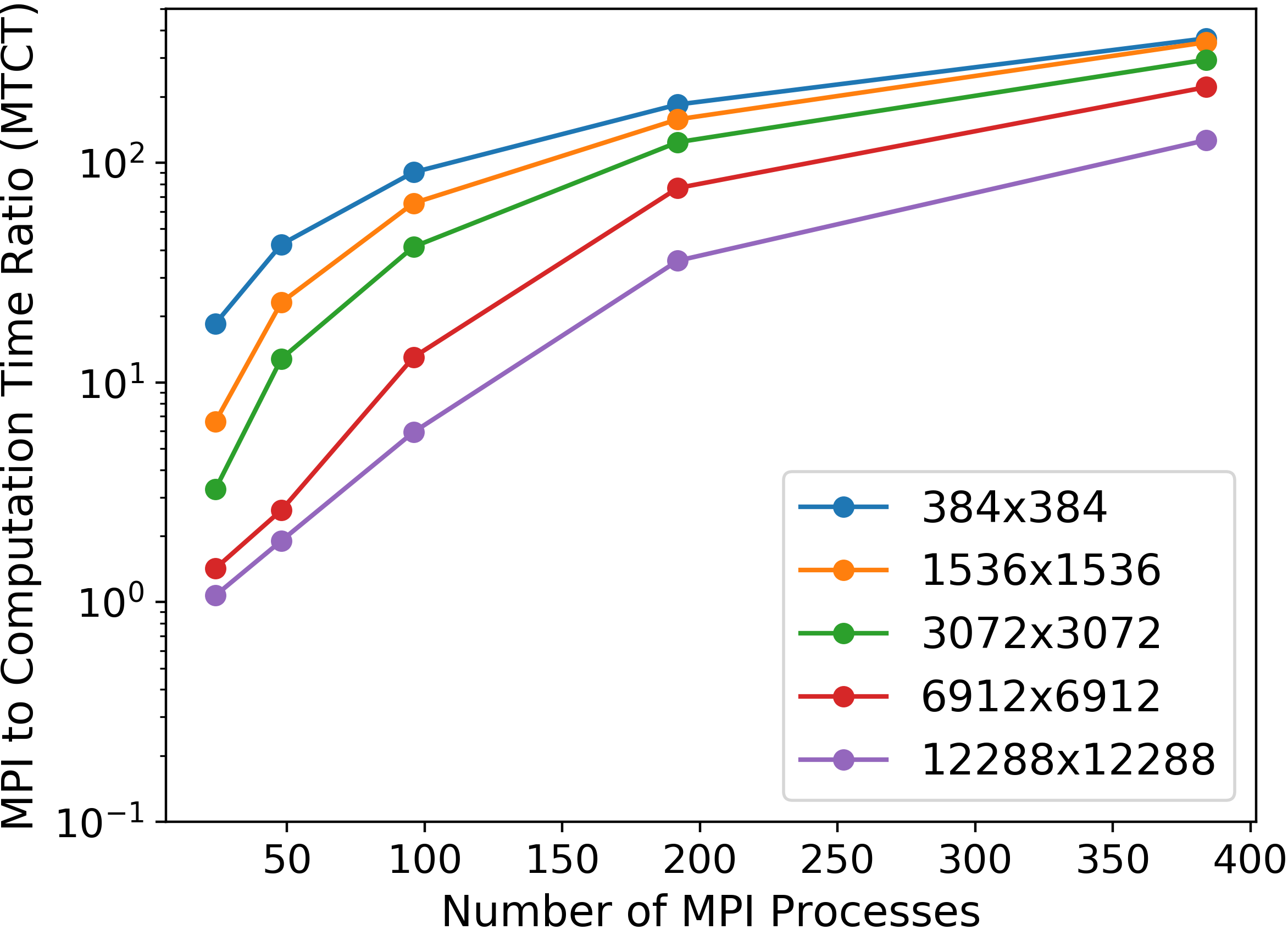}
\captionsetup{justification=centering}
\caption{MTCT ratio of a LU decomposition kernel for varying number of processes and matrix sizes.}
\label{fig:mtct}
\end{figure}

\subsection{Runtime Performance and Power Measurement}
\label{sec:perf_measurement}

An intuitive heuristic criterion for measuring the efficiency of a distributed MPI application is the time spent in MPI calls versus the time doing relevant computation, i.e., the MTCT ratio. The value of the metric depends upon the current number of processes and the input size as shown in Figure~\ref{fig:mtct}. In terms of performance-aware scheduling, the average value of this metric can be used for comparing and shrinking running malleable applications to start higher priority jobs waiting in the job queue (see Section~\ref{sec:perf_aware_scheduling}). For an iMPI application, the \texttt{SLURMD} daemons are responsible for collecting node-local performance measurements. This is done by aggregating process specific values from inside the iMPI library. To obtain the final metric values, a performance measurement handler was developed. The handler functions similarly to the one described in Section~\ref{sec:expand_shrink_jobs} and is responsible for obtaining and reducing the performance data from the appropriate \texttt{SLURMD} daemons. 

The mechanism for obtaining the power measurements is similar to the method developed for retrieving performance measurements. The power measurements for running applications are calculated from an automatically updated system file that stores energy values using RAPL counters. The average power value associated with the application is updated regularly. The latency for obtaining measurements scales with the number of processes due to the reduction operation and ranges from a hundred nanoseconds to two hundred microseconds~\cite{perf_measurement}.

\section{Adaptive Job Scheduling}
\label{sec:adaptive_job_scheduling}
We utilize the well-defined scheduling plugin API~\cite{schedpluginapi} provided by SLURM~\cite{yoo2003slurm} to implement our performance-aware and power-aware job scheduling strategies. In this section, we describe both the strategies in detail.

\subsection{Performance-aware scheduling of malleable jobs}
\label{sec:perf_aware_scheduling}

\begin{algorithm}[t]

\SetAlgoLined
\SetKwFunction{requestperfdata}{request\_perf\_data}
\SetKwFunction{shrinkjob}{shrink\_job}
\SetKwFunction{expandjob}{expand\_job}
\SetKwFunction{perfaware}{Perf\_Aware\_Schedule}
\SetKwProg{Pn}{Function}{:}{\KwRet}
\DontPrintSemicolon

\Pn{\perfaware}
{
    Obtain resource information from SLURM \;
    Obtain workload information from SLURM \;
    \While{requested resources available}
    {
        Start rigid and elastic jobs in priority order\;
    }
    \If{highest priority waiting job cannot be started}
    {
         \For{each running malleable job}
        {
            \If{job is adapting}
            {
                set $any\_job\_adapting$\;
            }
            \ElseIf{no performance data available for job}
            {
                \requestperfdata{$job$};
            }
        }
        \If{no malleable job is adapting and number of running malleable jobs $>$ 0} 
        {
            Analyze if running malleable jobs can be shrunk wrt decreasing MTCT ratios to start highest priority waiting job\;
            \If{enough nodes were found}{
                Shrink the selected malleable jobs. \;
                Start the highest priority waiting job. \;
            }
            \If{insufficient resources found or idle nodes available}{
                    Analyze running malleable jobs for expansion wrt increasing MTCT ratios. \;
                    Expand selected jobs. \;
            }
            
        }
    }
}

\caption{The ABS Performance-aware Scheduling function.}
\label{algo:batch_schded_algo}
\end{algorithm}

\algref{algo:batch_schded_algo} describes our methodology for performance-aware scheduling of rigid and malleable jobs. The \texttt{Perf\_Aware\_Schedule} function, which is responsible for launching rigid and elastic jobs, and dynamic reconfiguration decisions for running malleable applications, is event-triggered and runs on three events as described in Section~\ref{sec:overview_extensions}. Initially, the ABS obtains information about the current resources in the system, such as the  total number of compute nodes, the total number of idle nodes, etc, and also information about the user-submitted jobs in the elastic and rigid job queues (Line 2-3). The ABS tries to schedule and start as many rigid and malleable jobs as possible, depending upon the resource requirements and priorities (Line 4-6). The jobs are assigned priorities based on their arrival time as described in Section~\ref{sec:overview_extensions}.  All malleable jobs are launched with the resources specified in the \texttt{--nodes} parameter (see Listing~\ref{lst:sbatch_options}) and later shrunk or expanded depending upon the minimum and maximum number of nodes specified (see Section~\ref{sec:overview_extensions}). It is important to note that the \texttt{--nodes} parameter can be different from the minimum nodes required by the application. This gives more flexibility to the batch system in terms of expand/shrink operations, in contrast to previous strategies in which only the previously expanded jobs are considered for shrink operations~\cite{utrera2012job,sonmez2007scheduling}.

If a high priority job cannot be scheduled, due to the non-availability of required resources, then no lower-priority application is chosen for execution (Line 7). Following this, we iterate across the running malleable jobs in the system and check whether the job is currently adapting by using the \texttt{job\_state} variable (Line~8-9). If the job is found to be adapting then a flag variable is set (Line 10). Otherwise, a request is made for acquiring performance data for all running iMPI applications that do not have data associated with them, using the mechanism described in Section~\ref{sec:perf_measurement} (Line~13). The adaptation check is done to ensure consistency among compute nodes, and to assure that new reconfiguration decisions are generated after all adaptations have completed (see Section~\ref{sec:expand_shrink_jobs}).

If there are no jobs currently expanding and there is at least one running malleable job, then an expand/shrink decision phase is started (Line 16). Our scheduling methodology gives priority to waiting jobs and tries to start the highest priority job in the queue by shrinking the currently running malleable jobs. In contrast to previous approaches, the expand/shrink operations in the proposed algorithm consider the efficiency of the running application. In previous strategies either the resources are equally allocated/deallocated from the running jobs~\cite{sun2011fair} or the jobs are given priority for expand/shrink operations based upon their start time~\cite{utrera2012job, sonmez2007scheduling}. To start a higher priority waiting job, the algorithm tries to obtain the required resources by shrinking running jobs in the decreasing order of their MTCT ratios (Line 17). The ratio indicates the efficiency of the application as more MPI time for the same phase time is considered to be more inefficient. These shrink operations are mandatory and account for the minimum, maximum, and constraints on the number of nodes as specified by the user in the batch script (see Section~\ref{sec:overview_extensions}). For a valid value of the \texttt{--node-constraints} parameter, the shrink operation tries to reduce the number of nodes to the next lowest constraint value depending upon the remaining number of nodes required to start the waiting job. For instance, if the number of nodes allocated to a job are eight with \texttt{even} as the parameter specified in node constraints, and the remaining nodes required to start the waiting job are six, then the shrink operation will try to reduce the nodes to two. The selected running jobs undergo a shrink operation only if the required number of resources are obtained for the highest priority waiting job (Line 18-19). After the shrink operation completes, the waiting job is started (Line 20). If sufficient resources were not found or there are idle nodes in the system after the shrink operation, then the running jobs undergo an expansion phase. The running jobs are considered for expansion in the increasing order of their MTCT ratios (Line 23). This increases the resource utilization in the system with an increase in throughput as demonstrated in Section~\ref{sec:sched_results}. Similar to the shrink operation, the expansion algorithm accounts for the constraints on the number of nodes specified by the user. For a valid value of the \texttt{--node-constraints} parameter, the running job is allocated the highest constraint value, less than or equal to the current idle nodes. After distribution of idle nodes among the selected running jobs, they are expanded (Line 24).



\begin{algorithm}[t]
\SetAlgoLined
\SetKwFunction{poweraware}{Power\_Aware\_Schedule}
\SetKwProg{Fn}{Function}{:}{\KwRet}
\DontPrintSemicolon
\Fn{\poweraware}
{
    Obtain workload and resource information \;
    Update the power information for each running job\; 
    \If{no malleable job is adapting and number of running malleable jobs $>$ 0} 
        {
        \If{power corridor is broken}{
                \For{each waiting job $j$ in priority order}
                {
                    Calculate new resource distribution using LP with running jobs and waiting job $j$. \;
                    \If{feasible configuration found}
                        {
                            Redistribute resources. \;
                            Start the job $j$.\;
                            break;
                        }
                }
        }
        \If{power corridor is not broken}
        {
            Find the highest priority job(s) that satisfies the power constraints \;
            Start the selected jobs. \;
        }
    }
}
\caption{The ABS power-aware scheduling function.}
\label{algo:poweraware}
\end{algorithm}

An essential criterion for a malleable scheduling strategy is enabling fairness in dynamic reconfiguration decisions. While equipartitioning is a good strategy for achieving fairness in expand/shrink decisions, it is not always possible due to constraints on the number of nodes (see Section~\ref{sec:overview_extensions}). With our expand/shrink strategies we target overall system efficiency, while achieving  some fairness by resetting the MPI and phase time metric values for each job after completion of an adaptation operation. The metric values are also updated automatically if a significant change in the metric values is detected. Optimal selection of running jobs for expansion by using a performance prediction model can further improve the throughput of the system~\cite{martin2015enhancing} and is our interest for investigation in the future, but is out of scope for this work.

\subsection{Power-aware scheduling of malleable jobs}
\label{sec:power_aware_scheduling}
Our power-aware job scheduling strategy for dynamic power corridor management is described in \algref{algo:poweraware}. Similar to the performance aware strategy, the implemented scheduler function \texttt{Power\_Aware\_Schedule} executes on the three events as described in Section~\ref{sec:overview_extensions}. It is responsible for scheduling jobs that fit in the power budget and dynamic reconfiguration decisions to maintain/reinforce the system power within the power corridor.


At the beginning, the ABS obtains information about the current jobs and available resources in the system (Line 2). Initially, all jobs are launched according to the minimum and maximum power values per node mentioned by the user (See Section~\ref{sec:overview_extensions}). After launch, the  ABS updates the power values associated with each running job periodically (Line~3). Following this, if no malleable jobs are adapting and there are running malleable jobs, the ABS looks for a power corridor violation (Line 4, 5). If a power corridor violation is detected, we utilize a Linear Programming (LP) model shown in Equation~\ref{eq:heuristic} to enforce the power corridor.

\begin{footnotesize}
\begin{equation} \label{eq:heuristic}
\begin{split}
    Minimize:\\
    f(k_{idle}) &= k_{idle} * p_{idle} \\
    Subject \; To:\\
    l &\leq \sum_{i=1}^{K}k_{i}*p_{min}^{(i)} + k_{idle}*p_{idle} + m_{j}*p_{min}^{(j)} \\
    u &\geq \sum_{i=1}^{K}k_{i}*p_{max}^{(i)} + k_{idle}*p_{idle} + m_{j}*p_{max}^{(j)} \\
    k\_min_{i} &\leq k_i \leq k\_max_{i}, \; k_i \in \mathbb{N}  \setminus \{0\}, \; i = 1, \cdots, K \\
    0 &\leq k_{idle} < N, \; k_{idle} \in \mathbb{N} \\
\end{split}
\end{equation}
\end{footnotesize}
\newcommand{\textunderscript}[1]{$_{\text{#1}}$}

The objective of the LP model is to improve the system utilization, i.e, reduce the number of idle nodes in scheduling scenarios. The LP model uses running as well as waiting jobs to generate a new resource configuration that satisfies the power corridor. We iterate over the waiting jobs in priority order and use them along with the running jobs as an input to the LP model (Line 7). The first feasible resource redistribution configuration generated with the waiting job is utilized (Line~8). After this we expand/shrink the running jobs according to the configuration and start the selected waiting job (Line 9, 10). In Equation \ref{eq:heuristic}, \texttt{K} represents the number of running applications, \texttt{k\textunderscript{i}} represents the nodes required for the \texttt{job\textunderscript{i}}, \texttt{N} represents total number of nodes, \texttt{m\textunderscript{j}} represents the nodes required by the \texttt{job\textunderscript{j}} from the waiting queue in the order of priority, \texttt{k\_min\textunderscript{i}} and \texttt{k\_max\textunderscript{i}} represents the minimum and maximum values possible wrt to the \texttt{--node-constraints} of the malleable job, \texttt{P\textunderscript{min}} and \texttt{P\textunderscript{max}} represents the minimum and maximum power consumption of the job, and  \texttt{k\textunderscript{idle}} represents the number of idle nodes in the system consuming \texttt{P\textunderscript{idle}} power. Finally, \texttt{l} and \texttt{u} represents the lower and upper power boundary respectively.  If no power corridor violation is detected, we check the available power budget and start the appropriate jobs in priority order based on user data (Line 15-17).

\section{Experimental Results}
\label{sec:exp_results}

\begin{figure*}[t]
    \centering
    \begin{subfigure}{0.40\textwidth}
    \centering
        \includegraphics[width=0.86\linewidth]{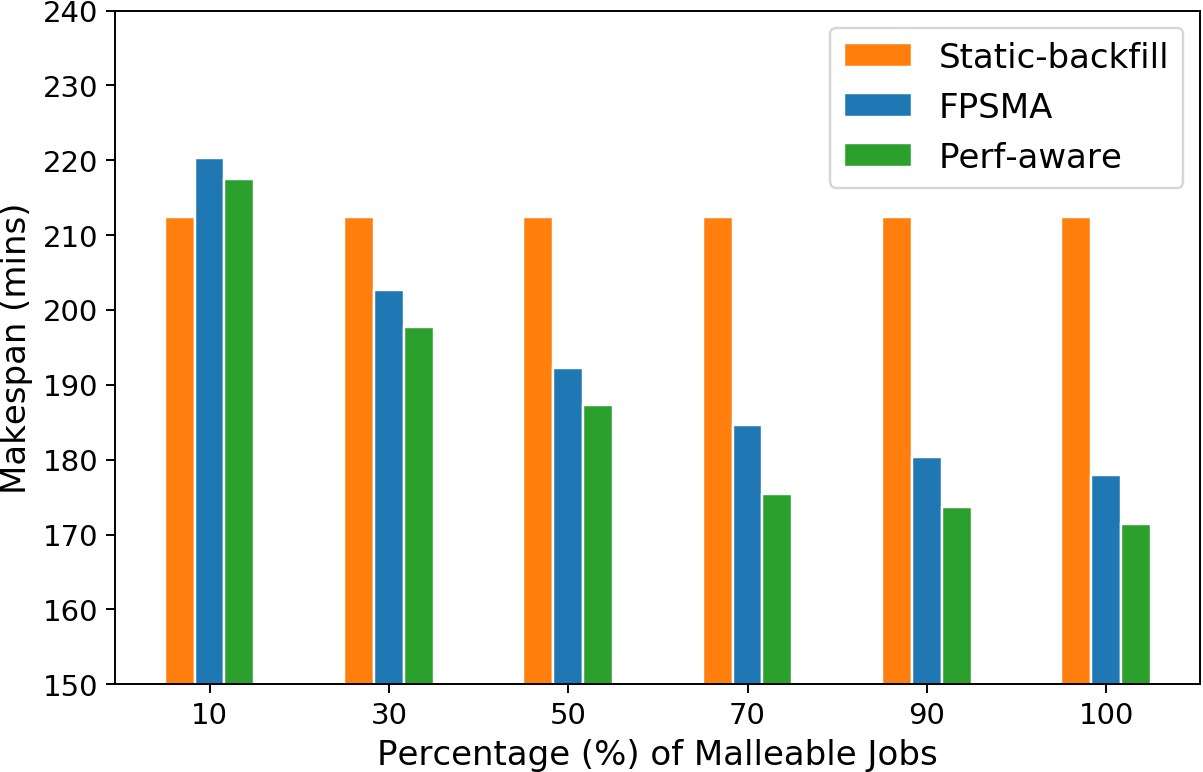}
        \caption{Makespan (Lower is better).}
        \label{fig:makespan}
    \end{subfigure}
    \begin{subfigure}{0.40\textwidth}
    \centering
        \includegraphics[width=0.86\linewidth]{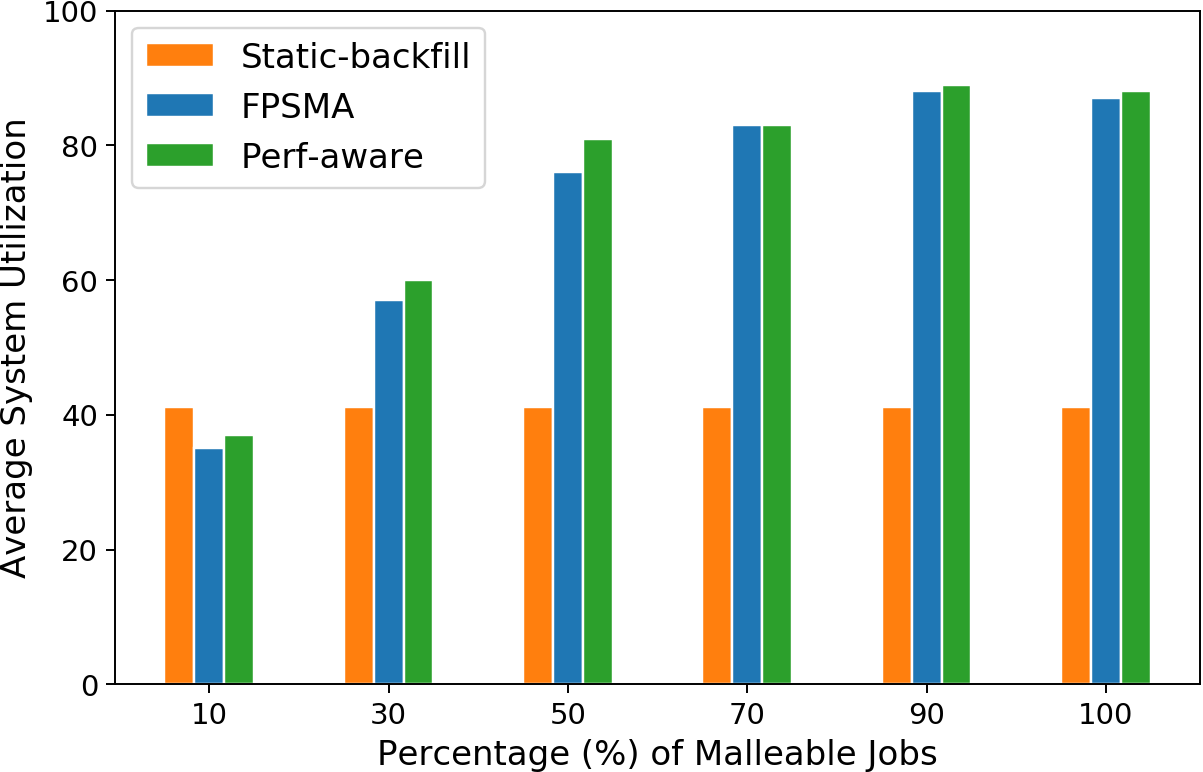}
        \caption{Average system utilization (Higher is better).}
        \label{fig:avgutil}    
    \end{subfigure}
    
    \begin{subfigure}{0.40\textwidth}
    \centering
        \includegraphics[width=0.86\linewidth]{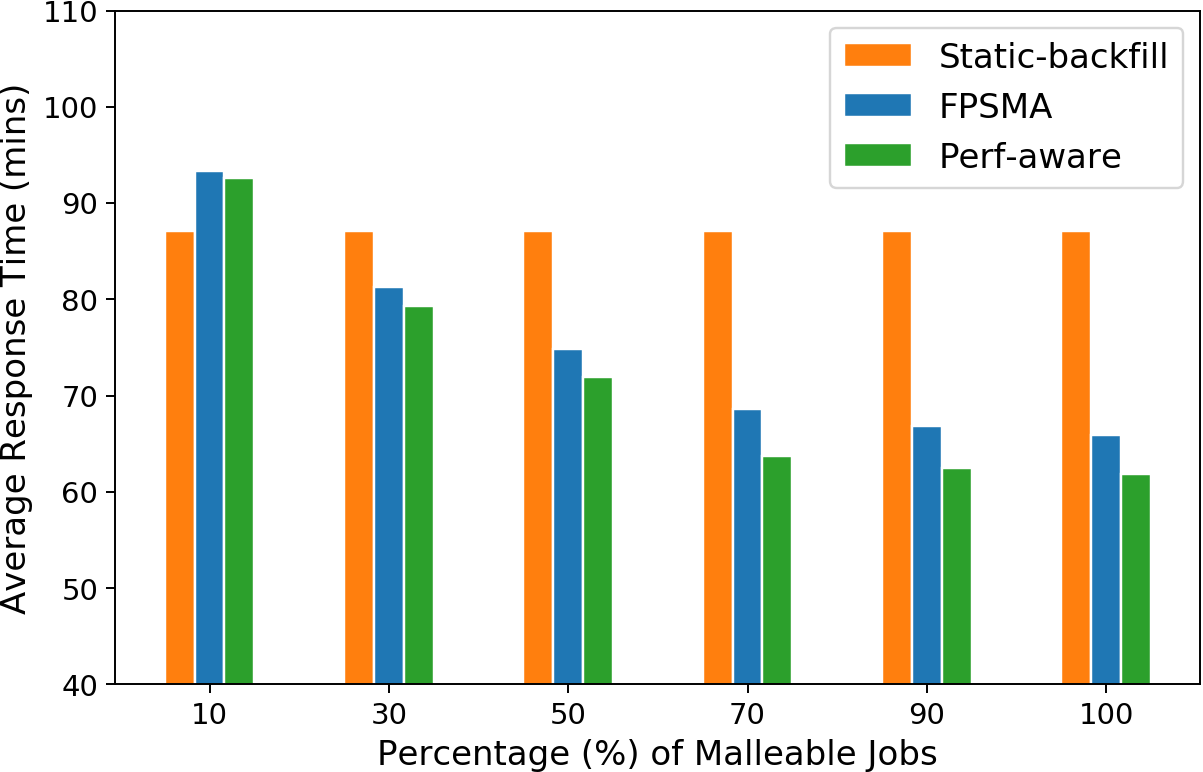}
        \caption{Average response time (Lower is better).}
        \label{fig:avgresp}
    \end{subfigure}
    \begin{subfigure}{0.40\textwidth}
    \centering
        \includegraphics[width=0.86\linewidth]{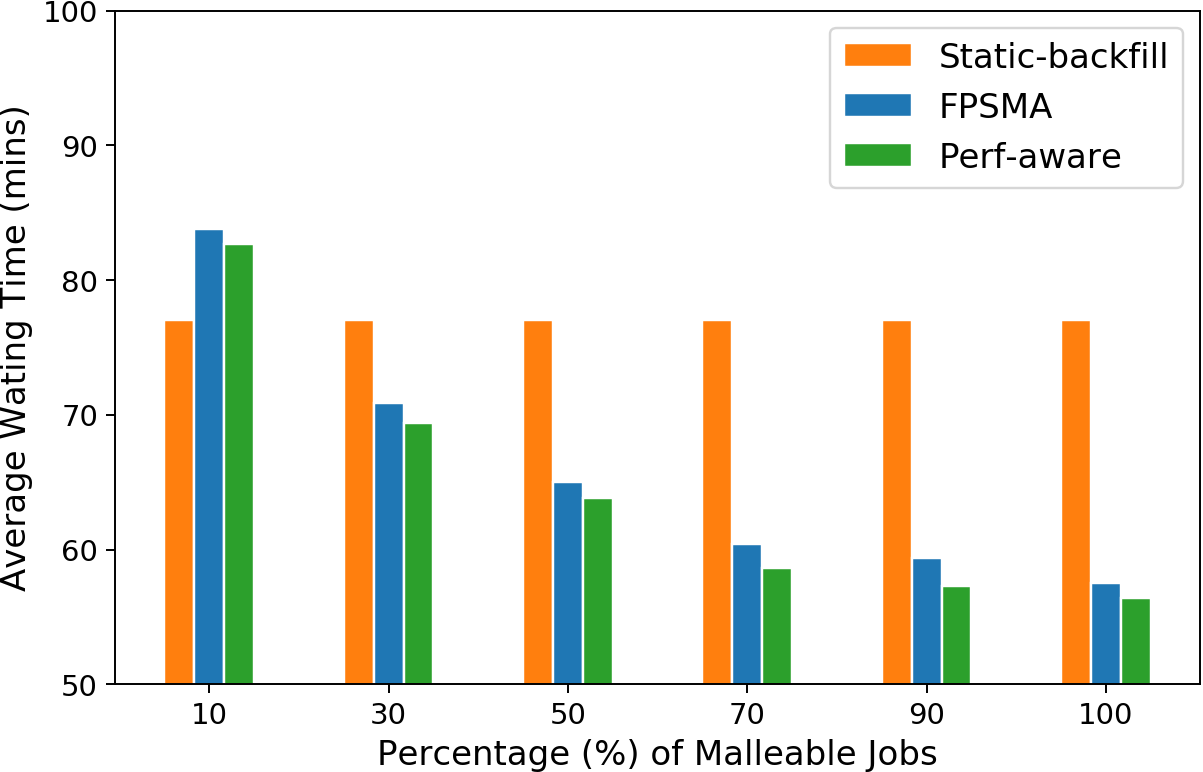}
        \caption{Average waiting time (Lower is better).}
        \label{fig:avgwait}
    \end{subfigure}
    \caption{Evaluation of the performance aware job scheduling strategy wrt system and user-centric for varying number of rigid and malleable jobs.}
    \label{fig_0}
\end{figure*}

    

In this section we evaluate the performance of our performance-aware batch scheduler wrt system and user-centric metrics and demonstrate dynamic power corridor management using our power-aware scheduling strategy.

\subsection{System Description}
\label{sec:exp_setup}
For analyzing the performance of the implemented batch system, we use the cluster SuperMUC-NG~\cite{lrz_supermuc} located at the Leibniz Supercomputing Center in Germany. SuperMUC-NG consists of eight islands comprising a total of $6480$ compute nodes based on the Intel Skylake-SP architecture. Each compute node has two sockets, comprising of two Intel Xeon Platinum 8174 processors, with 24 cores each and a total of 96GB main memory. The nominal operating core frequency for each core is 3.10 GHz. Hyper-Threading and Turbo Boost are disabled on the system. For evaluating the performance of our performance and power aware strategies, we simulate a virtual cluster with our extended SLURM as RJMS on 16 and 34 compute nodes of the SuperMUC-NG system respectively. One compute node is used as a login node for submitting the applications used for evaluation, and another compute node runs the ERS.  As a result, 14  and 32 compute nodes are available for running the applications on the virtual cluster respectively. 




\subsection{Performance-aware scheduler performance}
\label{sec:sched_results}

\begin{table}[t]
\caption{Synthetic workload characteristics for the modified ESP Benchmark~\cite{kramer2008percu}.}
\centering
\begin{adjustbox}{width=5.5cm,center}
      \begin{tabular}{|>{\centering\arraybackslash}m{0.5cm}|>{\centering\arraybackslash}m{1.5cm}| >{\centering\arraybackslash}m{0.65cm}|>{\centering\arraybackslash}m{1.25cm}|>{\centering\arraybackslash}m{1.25cm}|}
       \hline
        \textbf{Job Type} & \textbf{Fraction of System Size} & \textbf{Count} & \textbf{Static Execution Time [secs]} & \textbf{Constraints} \\ \hline 
        A & $0.03125$ & $75$ & 267 & - \\ \hline
        B & $0.06250$ & $9$ & 322 & \texttt{pof2} \\ \hline
        C & $0.50000$ & $3$ & 534 & - \\ \hline
        D & $0.25000$ & $3$ & 616 & \texttt{even}  \\ \hline
        E & $0.50000$ & $3$ & 315 & - \\ \hline
        F & $0.06250$ & $9$ & 1846 & \texttt{pof2} \\ \hline
        G & $0.12500$ & $6$ & 1334 & \texttt{even} \\ \hline
        H & $0.15625$ & $6$ & 1067 & \texttt{odd} \\ \hline
        I & $0.03125$ & $24$ & 1432 & - \\ \hline
        J & $0.06250$ & $24$ & 725 & \texttt{pof2} \\ \hline
        K & $0.09375$ & $15$ & 487 & - \\ \hline
        L & $0.12500$ & $36$ & 366 & \texttt{even} \\ \hline
        M & $0.25000$ & $15$ & 187 & - \\ \hline
        Z & $1$ & $2$ & 100 & - \\ \hline
   \end{tabular}
    
 \end{adjustbox}
 \label{tab:esp_benchmark}
\end{table}

To evaluate and analyze the performance of our scheduling strategy, we adopt and modify the Effective System Performance (ESP)~\cite{kramer2008percu} benchmark. The ESP benchmark provides a quantitative evaluation of the performance of an RJMS software and is an efficient method for comparing different scheduling strategies~\cite{georgiou2010contributions, prabhakaran2015batch}. It consists of 230 jobs derived from 14 job types with each job type having a fixed unique execution time and running the same synthetic application. The job types along with their instance counts, fraction of total system size, and target runtimes are shown in Table~\ref{tab:esp_benchmark}. We replace the synthetic application with a Tsunami simulation\footnote{https://github.com/mohellen/eSamoa}~\cite{mo2017large} modeled using the 2-D shallow water wave equation and programmed using iMPI. The Tsunami simulation is an example of a real world scientific application implemented using a framework for dynamically adaptive meshes called $sam(oa)^{2}$. The framework is responsible for grid refinement and load balancing at each time step of the simulation. We achieve different target runtimes for each job type (see Table~\ref{tab:esp_benchmark}) by changing the grid resolution and total simulation time. Furthermore, we add constraints on the number of nodes for dynamic reconfiguration decisions on seven job types (see Section~\ref{sec:overview_extensions}). We utilize 34 compute nodes to conform with the  requirements of the ESP benchmark.

We compare the performance of our performance aware scheduling strategy with static backfilling and Favour Previously Started Malleable applications (FPSMA) first~\cite{sonmez2007scheduling} strategies. In the FPSMA strategy, jobs are considered for expand/shrink operations in the increasing/decreasing order of their start times. We quantify the performance using four metrics, i.e., makespan, average system utilization, average waiting and response times. Makespan represents the difference between the last job end time and arrival time of the first job. Average system utilization describes the fraction of the total system utilized during the entire workload execution. Average waiting time represents the difference between start and submission times, averaged across all jobs. Average response time is the sum of waiting time and runtime averaged for all jobs. 

Figure~\ref{fig_0} shows the values of the metrics for the three strategies for varying number of rigid and malleable jobs. All jobs in the ESP benchmark are submitted to the ABS in a random order determined by a pseudo-random generator, with the inter-arrival time between jobs being fixed to 30 seconds. The order of submission of jobs is the same for all strategies and all scenarios. Malleable jobs are selected by using a pseudo-random generator with a fixed seed and are identical for both FPSMA and performance aware strategies. For all malleable jobs the \texttt{--nodes} parameter is set to the corresponding system size with minimum and maximum number of nodes set to lowest and highest value of the specified node constraint respectively. In the case of malleable jobs with no constraints the minimum and maximum nodes are set to one and 32. The metric values for the static backfilling strategy are obtained by running the default \texttt{backfill} scheduling plugin in SLURM with default scheduling parameters. In this case, no expand/shrink operations are considered.

For the scenario with \texttt{100\%} malleable jobs the implemented performance aware scheduling strategy obtains makespan, average response and waiting time values $19.3\%$, $29.0\%$, and $26.8\%$ lower than the static backfilling strategy (see Figures~\ref{fig:makespan}, \ref{fig:avgresp}, \ref{fig:avgwait}). In comparison to the malleable FPSMA strategy the metric values obtained are $4.0\%$, $6.1\%$, and $2.0\%$ lower respectively. The performance aware strategy performs better than the backfilling approach in all cases except for the scenario with $10\%$ malleable jobs. In this case, both the FPSMA and performance aware strategies have a higher value for the three metrics. This can be attributed to two reasons. Firstly, constraints on the number of nodes wrt dynamic reconfiguration decisions for different job types. Secondly, since both malleable scheduling strategies follow a FCFS scheduling policy, the rigid backfilling strategy benefits from efficient selection of jobs from the submitted job queue that can be started immediately without violating the resource reservations of the highest priority waiting jobs.

Figure~\ref{fig:avgutil} shows the comparison between average system utilization for the three strategies. The malleable scheduling strategies outperform the backfilling approach for all cases except for the scenario with $10\%$ malleable jobs. While the FPSMA and the performance aware scheduling strategies have a similar average system utilization, the implemented performance aware strategy has a higher throughput. This can be attributed to the inefficient selection of jobs for expand/shrink operations based on their start times in the FPSMA strategy as compared to the current value of the \texttt{MTCT} ratio in our performance aware approach. This can also be explained based upon the nature of our scientific application which utilizes AMR~\cite{plewa2005adaptive}.  The simulation undergoes a grid refinement process periodically leading to a change in it's communication and computational requirements. Therefore, a scheduling strategy based on a heuristic criterion that represents the efficiency of the application leads to a better performance. We also evaluated the performance of other malleable job scheduling strategies which are described in~\cite{chadhaadaptive}.




\subsection{Dynamic Power Corridor management}
\label{sec:dyn_power_corridor}
To analyze our power-aware scheduling strategy we utilize two iMPI applications with different power requirements. The first application repeatedly calculates the value of \texttt{PI} and the second solves the 2D heat equation using the Jacobi iteration. Our workload consists of 20 jobs with equal distribution among the two applications. The nodes (\texttt{--nodes}) required by the applications range from 1 to 4. The average power consumed by the Pi and the heat applications are 170 and 250 watts per node respectively. The jobs are submitted to the ABS alternatively with an inter-arrival time of two seconds. The minimum and maximum nodes for all jobs is set to one and 14.

\begin{table}[t]
\caption{Comparison of metric values for different scheduling scenarios.}
\centering
 \begin{adjustbox}{width=4.5cm,center}
      \begin{tabular}{|>{\centering\arraybackslash}m{1cm}|>{\centering\arraybackslash}m{2cm}| >{\centering\arraybackslash}m{1.2cm}|>{\centering\arraybackslash}m{2cm}|}
       \hline
        \textbf{Scenario} & \textbf{Power corridor violations} & \textbf{Makespan (mins)} \\ \hline 
        1 & $6$ & $13.75$  \\ \hline
        2 & $2$ & $14.25$  \\ \hline
        3 & $0$ & $14.00$  \\ \hline

   \end{tabular}
    
  \end{adjustbox}
 \label{tab:power_aware}
\end{table}
\begin{figure}[t]
\centering
\includegraphics[width=0.68\columnwidth]{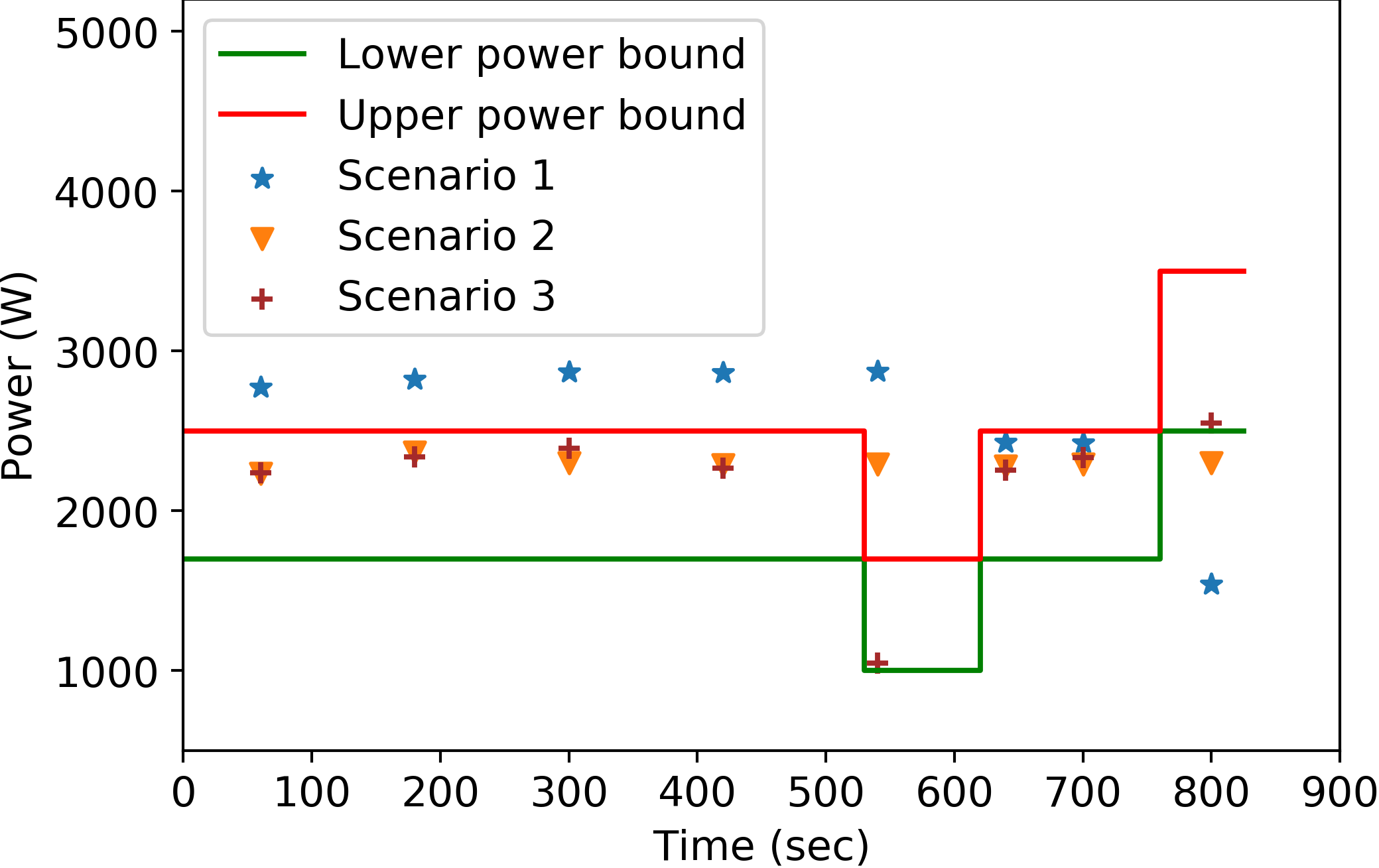}
\captionsetup{justification=centering}
\caption{Average system power usage and power corridor violations for different scheduling scenarios.}
\label{fig:power}
\end{figure}

To compare and evaluate the benefits of our strategy we consider three scenarios. In scenario 1, we use the static \texttt{backfill} scheduler in SLURM and no redistribution of resources. For scenarios 2 and 3, we utilize the LP model to generate node redistributions and enforce the power corridor. However, in scenario 2, only running jobs are considered for the redistribution~\cite{john2020invasive}, while in scenario 3 waiting jobs are also taken into account (See~\algref{algo:poweraware}). 
The order of submission of jobs for all the scenarios is the same and all jobs are considered to be malleable in scenarios 2 and 3. For all scenarios, we dynamically change the values of the power corridor at fixed times. Initially, the lower and upper power bound is set to 1700 and 2500 respectively and is reduced to 1000 and 1700 Watts. Later, the power corridor value is increased to 2500 and 3500 Watts. For obtaining the minimum value of the lower bound, we calculated the power consumed by the idle nodes. The average idle power consumed by a compute node on our system is 71 Watts. The maximum value of the upper bound is obtained by assuming that the heat simulation was running on 14 nodes. 

The number of power corridor violations and the makespan obtained in the scheduling scenarios are shown in \tabref{tab:power_aware}. Number of power corridor violations represents whether the system was able to reconfigure the resources to restore the power corridor. Although the backfilling strategy leads to better makespan than the malleable strategies, we observe six power corridor violations for it. This can be attributed to no resource redistributions. The comparison between scenarios 2 and 3 can be better understood from Figure \ref{fig:power}. Initially, both scenarios 2 and 3 were able to maintain the power corridor through resource reconfigurations. When the power corridor was decreased dynamically, scenario 3 was able to maintain the power corridor by generating a new feasible resource reconfiguration using the LP model and a waiting job. On the other hand, for scenario 2 no such feasible resource reconfiguration was found for the running jobs. 
A similar situation occured when the power corridor was changed again. In our implementation, we tested the combinations of length one and two of the number of jobs in waiting queue as an input to the LP model. In all cases, we found a feasible solution with a single waiting job. We observed an average overhead of 20 milliseconds for obtaining resource configuration from our LP model. The number of combinations to be tested with the LP model can be increased, however it leads to a higher overhead.

\subsection{Analyzing Overhead}
\label{sec:anal_overhead}
\begin{figure}[t]
\centering
\includegraphics[width=0.68\columnwidth]{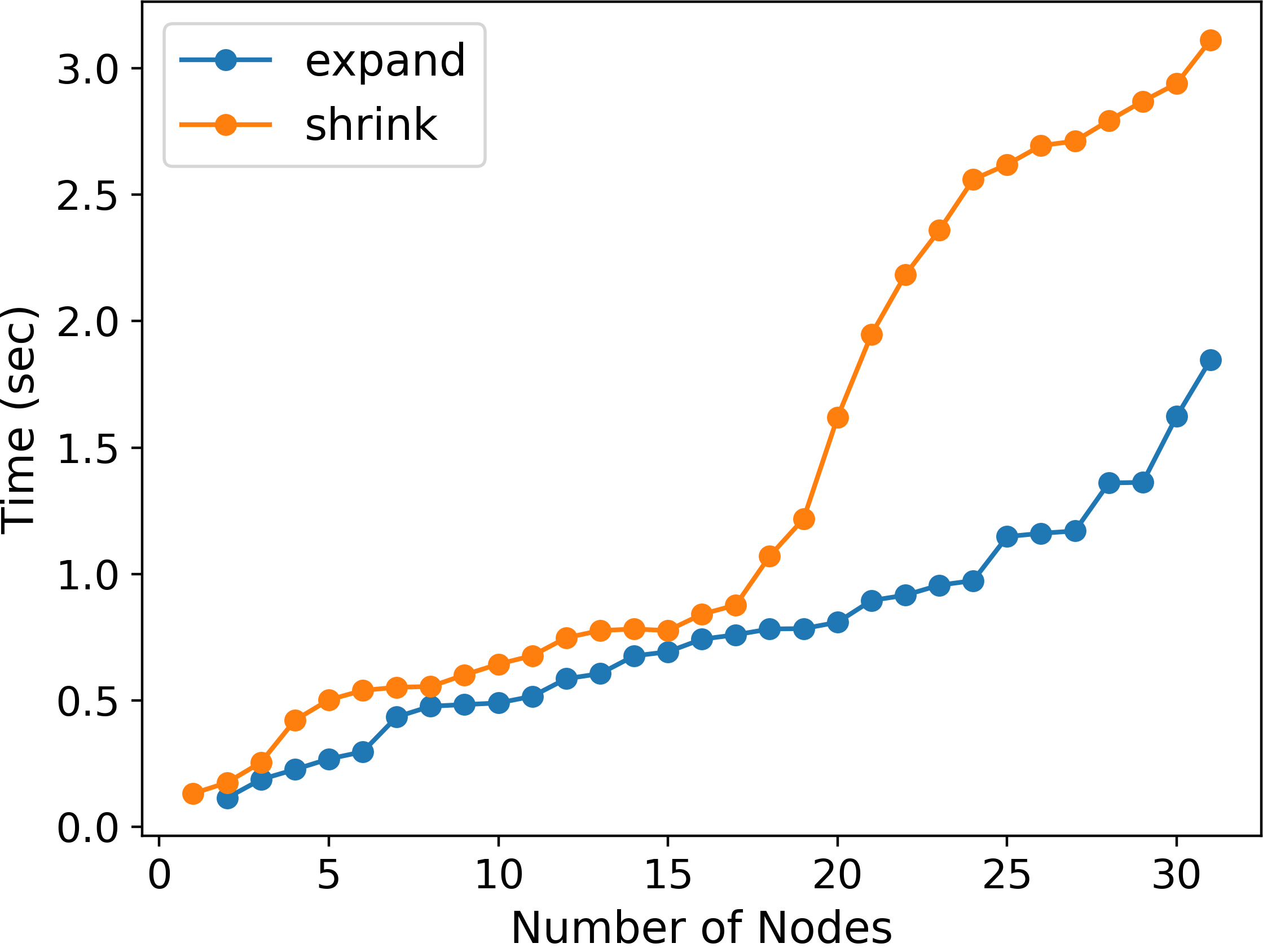}
\captionsetup{justification=centering}
\caption{Latency for expand and shrink operations for a job running initially on 1 and 32
nodes respectively.}
\label{fig:latency}
\end{figure}
To quantify the overhead for expand/shrink operations in our infrastructure we measure the time between the start and end of an adaptation window using a synthetic iMPI application (see Section~\ref{sec:impi}). The values are obtained after averaging the results over five runs. For expansion, the job initially starts from one compute node and is grown periodically by one node, upto 32 nodes. In the case of reduction, the job initially starts from 32 nodes and is shrunk periodically by one node. Each expand/shrink operation involves the addition/removal of $48$ processes. The obtained latency values for these operations is shown in Figure~\ref{fig:latency}. The total time required for expansion operation increases with the increasing number of nodes which can be attributed to the latency in launching the new processes and communication with a greater number of nodes. Similar to expansion, the latency for shrink operations also increases with higher number of nodes. However, the observed overhead is higher as compared to expansion operations since it depends upon the time required for preexisting tasks on preexisting nodes to complete. Each task must send a notification to
the ERS after completion. We observe a maximum overhead of $3.1$ and $1.8$ seconds for expand/shrink operations respectively. The latency values do not include any data redistribution overhead. In the future, we plan to investigate the use of scalable process management interface (PMIx)~\cite{CASTAIN20189} in our infrastructure to further reduce overhead for expand/shrink operations.

\section{Conclusion \& Future Work}
\label{sec:conclusion}

In this paper, we extended the SLURM batch system for dynamic resource-aware batch scheduling of malleable appplications written using a new adaptive parallel paradigm called Invasive MPI. We proposed and implemented two scheduling strategies in SLURM for performance-aware and power-aware scheduling of jobs. Through experiments on a production HPC system, we demonstrated an improvement in system and user-centric metrics for our performance-aware strategy as compared to the commonly used static-backfill and FPSMA policies. We showed that our power-aware scheduling strategy is able to maintain the system power consumption within the power corridor when the upper and lower bounds are changed dynamically. In the future, we plan to explore the usage of frequency scaling and power capping for a hybrid power-aware scheduling strategy. Furthermore, we plan to extend our infrastructure to support malleability at thread level granularity.

\section{Acknowledgment}
\label{sec:ack}
The research leading to these results was funded by the Deutsche Forschungsgemeinschaft 
(DFG, German Research Foundation)-Projektnummer 146371743-TRR 89: Invasive Computing.


\bibliographystyle{IEEEtran}

\bibliography{parallelpgm}


\end{document}